\pdfoutput=1
\documentclass[anonymous=false,
               format=acmsmall,
               review=false,
               screen=true,
               nonacm=true]{acmart}

\usepackage{amsmath,amsthm}
\allowdisplaybreaks
\newtheorem{theorem}{Theorem}

\theoremstyle{definition}
\newtheorem{definition}[theorem]{Definition}
\newtheorem{example}[theorem]{Example}

\usepackage{amssymb,stmaryrd}

\usepackage{mathtools}

\usepackage{calc}
\usepackage{tikz}
\usetikzlibrary{shapes,positioning,calc,fit,backgrounds,decorations.pathmorphing}
\tikzset{var/.style={draw,circle,fill=white,inner sep=1.5pt,minimum size=8pt}}
\tikzset{ext/.style={var,fill=black,text=white}}
\tikzset{fac/.style={draw,rectangle}}
\tikzset{subgraph/.style={draw,cloud}}
\tikzset{plate/.style={draw,rectangle,rounded corners}}
\tikzset{every picture/.style={baseline=-2.5pt}}
\tikzset{every label/.style={font={\footnotesize}}}
\tikzset{node distance=0.5cm and 0.5cm}

\usepackage{comment}
\usepackage{xcolor}
\usepackage{verbatim}

\newcommand{\asst}{\xi}
\newcommand{\Asst}{\Xi}

\newcommand{\arity}{\mathit{arity}}
\newcommand{\att}{\mathit{att}}
\newcommand{\ext}{\mathit{ext}}
\newcommand{\lab}{\mathit{lab}}

\newcommand{\alt}{\mathrel{|}}

\newcommand{\bnf}{\mathrel{::=}}
\newcommand{\bnfalt}{\mathrel{\makebox[\widthof{$\bnf$}]{$|$}}}

\newcommand{\rv}[1]{\mathbf{#1}}
\newcommand{\nt}[1]{\mathsf{#1}} 

\newcommand{\shrink}[1]{\par\noindent\scalebox{0.8}{\begin{minipage}{1.25\textwidth}\par\vspace{\abovedisplayskip}#1\par\end{minipage}}\par\vspace{\belowdisplayskip}}

\newcommand{\inl}[1]{\ensuremath{\textit{inl\/}(#1)}}
\newcommand{\inr}[1]{\ensuremath{\textit{inr\/}(#1)}}

\acmConference[PROBPROG]{Conference on Probabilistic Programming}{2020}{Online}

\title[Translating Recursive Probabilistic Programs to Factor Graph Grammars]{Translating Recursive Probabilistic Programs \linebreak[2] to Factor Graph Grammars}

\author{David Chiang}
\orcid{0000-0002-0435-4864}
\affiliation{
  \department{Department of Computer Science and Engineering}
  \institution{University of Notre Dame}
  \city{Notre Dame}
  \state{IN}
  \country{USA}
}
\email{dchiang@nd.edu}

\author{Chung-chieh Shan}
\orcid{0000-0002-0339-6405}
\affiliation{
  \department{Department of Computer Science}
  \institution{Indiana University}
  \city{Bloomington}
  \state{IN}
  \country{USA}
}
\email{ccshan@indiana.edu}

\begin{document}

\begin{abstract}
    It is natural for probabilistic programs to use conditionals
    to express alternative substructures in models, and
    loops (recursion) to express repeated substructures in models.
    Thus, probabilistic programs with conditionals and recursion
    motivate ongoing interest in efficient and general inference.
    A~\emph{factor graph grammar} (FGG) generates a set of factor graphs
    that do not all need to be enumerated in order to perform inference.
    We provide a semantics-preserving translation from first-order
    probabilistic programs with conditionals and recursion to FGGs.
\end{abstract}

\maketitle

\section{Introduction}

Probabilistic models often contain repeated substructures, such as the time steps of a hidden Markov model, as well as alternative substructures, such as different grammar productions that generate the same string.
Loops (recursion) and conditionals naturally and compactly represent such substructures in probabilistic programming.
However, efficient and general inference on these representations remains a longstanding challenge.

\emph{Factor graph grammars} (FGGs) \citep{chiang+riley:2020} are hyperedge replacement grammars (HRGs) \citep{bauderon+courcelle:1987,habel+kreowski:1987,drewes+:1997} for generating sets of factor graphs. They have recently been proposed as a unified formalism that can describe both repeated and alternative substructures.
Moreover, it turns out that inference can be performed on an FGG (by variable elimination) without enumerating all the factor graphs generated.

This paper explores the relationship of this formalism to probabilistic programming languages (PPLs). When random variables are limited to finite domains, what kinds of models can and can't be expressed using FGGs? We answer this question by defining a simple PPL and giving a translation from this language to FGGs.
This translation extends that of \citet{vandemeent+:2018}, which translates a PPL without higher-order functions or recursion to a factor graph.
Our source language does allow recursion, and our translation uses the added power of FGGs to handle recursion and conditionals cleanly.

\section{Definitions}

To establish notation, we briefly define hypergraphs, factor graphs, HRGs, and finally FGGs.

Fix a finite set $L$ of \emph{edge labels}, and a function $\arity : L \rightarrow \mathbb{Z}_{\geq 0}$.

\begin{definition}
A \emph{(hyper)graph} is a tuple $(V, E, \lab, \att, \ext)$, where
\begin{itemize}
\item $V$ names a finite set of \emph{nodes}.
\item $E$ names a finite set of \emph{(hyper)edges}.
\item $\lab: E \rightarrow L$ assigns to each edge a label.
\item $\att: E \rightarrow V^\ast$ assigns to each edge~$e$ a sequence of $\arity(\lab(e))$ \emph{attachment} nodes. 
\item $\ext \in V^\ast$ is a sequence of zero or more \emph{external} nodes.
\end{itemize}
\end{definition}


\begin{definition} \label{def:factorgraph}
A \emph{factor graph} \citep{kschischang+:2001} is a hypergraph $(V, E, \lab, \att, \ext)$ together with a set~$D$ and a function~$F$:
\begin{itemize}
\item $D$ is the set of possible values that each node can take. For simplicity, we use a single $D$ for all nodes, unlike in the previous definition \citep{chiang+riley:2020}.
\item $F$ maps each edge label~$\ell$ to a function $F(\ell) : D^{\arity(\ell)} \rightarrow \mathbb{R}_{\geq 0}$. For brevity, we write $F(e)$ for $F(\lab(e))$.
An edge $e$ together with its function $F(e)$ is called a \emph{factor}.
\end{itemize}
\end{definition}

\begin{example}
  \label{eg:fg_pcfg}
  Below is a factor graph for trees of a certain shape generated by a PCFG with start symbol $S$. Variables $\rv{N}$ range over nonterminal symbols and $\rv{W}$ over terminal symbols.
  \shrink
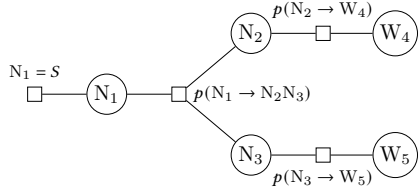
{\begin{center}
      \begin{tikzpicture}[x=1.2cm]
        \node[var] (n1) at (0,0) { $\rv{N}_1$ };
        \node[var] (n2) at (2,1) { $\rv{N}_2$ };
        \node[var] (n3) at (2,-1) { $\rv{N}_3$ };
        \node[var] (w4) at (4,1) { $\rv{W}_4$ };
        \node[var] (w5) at (4,-1) { $\rv{W}_5$ };
        \node[fac,label=above:{$\rv{N}_1 = S$}] at (-1,0) {} edge (n1);
        \node[fac,label=right:{$p(\rv{N}_1\rightarrow \rv{N}_2 \rv{N}_3)$}] at (1,0) {} edge (n1) edge (n2) edge (n3);
        \node[fac,label=above:{$p(\rv{N}_2\rightarrow \rv{W}_4)$}] at (3,1) {} edge (n2) edge (w4);
        \node[fac,label=below:{$p(\rv{N}_3\rightarrow \rv{W}_5)$}] at (3,-1) {} edge (n3) edge (w5);
      \end{tikzpicture}
  \end{center}}
\end{example}

We draw a node as a circle with its name inside, and an edge as a square with lines (called \emph{tentacles}) to its attachment nodes. To reduce clutter, we don't show the ordering of tentacles.
We draw a factor $e$ as a small square with $F(e)$ next to it, as an expression in terms of its attachment nodes' names. A~Boolean expression has value $1$ if true and $0$ if false.

\begin{definition}
An \emph{assignment} of
a factor graph $H = (V, E, \lab, \att, \ext)$ is a mapping $\asst : V \rightarrow D$ from nodes to values. We write $\Asst_H$ for the set of all assignments of $H$.
The \emph{weight} of an assignment~$\asst$~is
\begin{equation}
    w_H(\asst) = \prod_{e\in E} F(e)\bigl(\asst(\att(e)_1), \ldots, \asst(\att(e)_{|\att(e)|})\bigr).
\end{equation}
We also define the marginal distribution over assignments to external nodes, $\asst : \ext \rightarrow D$:
\begin{equation}
  w_{\ext}(\asst) = \sum_{\mathclap{\substack{\asst' \in \Asst_H \\ \left.\asst'\right|_{\ext} = \asst}}} w_H(\asst').
\end{equation}
\end{definition}

\begin{definition} \label{def:hrg}
A \emph{hyperedge replacement graph grammar} (HRG) is a tuple $(N, T, P, S)$, where
\begin{itemize}
\item $N \subseteq L$ is a finite set of \emph{nonterminal symbols}.
\item $T \subseteq L$ is a finite set of \emph{terminal symbols}, disjoint from~$N$.
\item $P$ is a finite set of \emph{productions} (or \emph{rules}) of the form $(X \rightarrow R)$, where 
    $X \in N$, and
    $R$ is a hypergraph with edge labels from $N \cup T$
    and exactly $\arity(X)$ external nodes.
\item $S \in N$ is a distinguished \emph{start nonterminal symbol}.
\end{itemize}
\end{definition}
Most definitions require $\arity(S) = 0$, including the previous definition of FGGs \citep{chiang+riley:2020}, but here we relax this requirement, to allow the graphs generated by an HRG to have external nodes.

\begin{definition}
If $G$ is an HRG and $X$ is a nonterminal symbol, define the set of \emph{$X$-derivation trees} of~$G$, $\mathcal{D}_G(X)$, to be the smallest set containing all pairs $(R, s)$ such that $(X \rightarrow R) \in G$ and $s$ is a mapping from each nonterminal edge~$e$ in~$R$ to a $\lab(e)$-derivation tree of~$G$.
Define the set of derivation trees of $G$ to be $\mathcal{D}_G = \mathcal{D}_G(S)$.

An $X$-derivation tree $T = (R, s)$ \emph{yields} a graph with the same $\arity(X)$ external nodes as~$R$.
We define this graph by induction on~$T$ as follows.
For each $e \in R$, let $H_e$ be the graph yielded by $s(e)$. Replace $e$ with $H_e$, identifying each attachment node $\att(e)_i$ of $e$ with the $i$th external node of $H_e$; the resulting node is external iff $\att(e)_i$ was.
\end{definition}

\begin{definition}
An HRG for factor graphs, or a \emph{factor graph grammar} (FGG) for short, is an HRG together with a set~$D$ and and a function~$F$, as in the definition of factor graphs (Definition \ref{def:factorgraph}), except that $F$ is defined on terminal edge labels only.
\end{definition}

\begin{example}
\label{eg:fgg_pcfg}
Below is an FGG for derivations of a PCFG in Chomsky normal form. The start symbol of the FGG is $\nt{S'}$.
\shrink
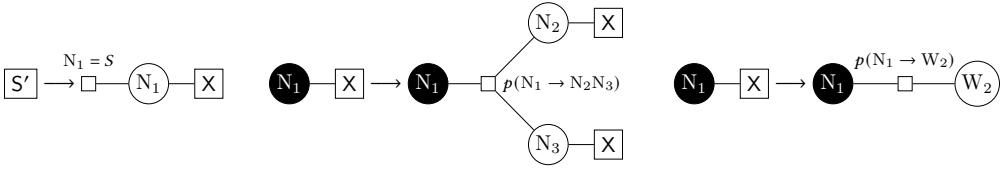
{%
\begin{align*}
\begin{tikzpicture} 
\node[fac] at (1,0) { $\nt{S'}$ };
\end{tikzpicture} 
&\longrightarrow 
\begin{tikzpicture} 
\node[var] (n) at (1,0) { $\rv{N}_1$ }; 
\node[fac,label=above:{$\mathclap{\rv{N}_1 = S}$}] at (0,0) {} edge (n);
\node[fac] at (2,0) {$\nt{X}$} edge (n); 
\end{tikzpicture} &
\quad
\begin{tikzpicture} 
\node[ext](x) at (0,0) {$\rv{N}_1$};
\node[fac] at (1,0) { $\nt{X}$ } edge (x); 
\end{tikzpicture} 
&\longrightarrow 
\begin{tikzpicture}
\node[ext] (n) at (0,0) { $\rv{N}_1$ };
\node[var] (n1) at (2,1) {$\rv{N}_2$};
\node[var] (n2) at (2,-1) {$\rv{N}_3$};
\node[fac,label=right:{$p(\rv{N}_1\rightarrow \rv{N}_2 \rv{N}_3)$}] at (1,0) {} edge (n) edge (n1) edge (n2);
\node[fac] at (3,1) {$\nt{X}$} edge (n1);
\node[fac] at (3,-1) {$\nt{X}$} edge (n2);
\end{tikzpicture} &
\quad
\begin{tikzpicture}
  \node[ext](x) at (0,0) {$\rv{N}_1$};
  \node[fac] at (1,0) { $\nt{X}$ } edge (x);
\end{tikzpicture}
&\longrightarrow
\begin{tikzpicture}[x=1.2cm]
  \node[ext] (n) at (0,0) { $\rv{N}_1$ };
  \node[var] (n1) at (2,0) { $\rv{W}_2$ };
  \node[fac,label=above:{$p(\rv{N}_1 \rightarrow \rv{W}_2)$}] at (1,0) {} edge (n) edge (n1);
\end{tikzpicture}
\end{align*}}
This FGG generates an infinite number of factor graphs, including the one in Example~\ref{eg:fg_pcfg}, whose derivation tree looks like:
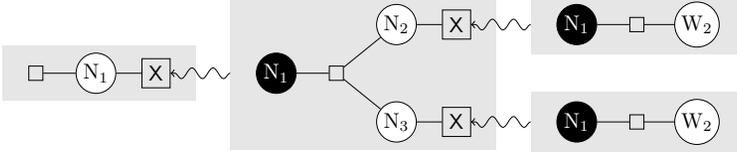
\shrink{\begin{center}
  \begin{tikzpicture}
    \tikzset{replace/.style={->,decorate,decoration=snake}}
    \begin{scope}
      \node[var](n1) at (1,0) {$\rv{N}_1$};
      \node[fac](f0) at (0,0) {} edge (n1);
      \node[fac](f1) at (2,0) { $\nt{X}$ } edge (n1);
      \begin{scope}[on background layer]
        \node[fit=(f0)(n1)(f1),fill=gray!20,inner xsep=12pt] {};
      \end{scope}
    \end{scope}
    \begin{scope}[xshift=4cm,y=0.8cm]
      \node[ext] (n) at (0,0) { $\rv{N}_1$ };
      \node[var] (n1) at (2,1) {$\rv{N}_2$};
      \node[var] (n2) at (2,-1) {$\rv{N}_3$};
      \node[fac] at (1,0) {} edge (n) edge (n1) edge (n2);
      \node[fac] (f4) at (3,1) {$\nt{X}$} edge (n1);
      \node[fac] (f5) at (3,-1) {$\nt{X}$} edge (n2);
      \begin{scope}[on background layer]
        \node[fit=(n)(n1)(n2)(f4)(f5),fill=gray!20,inner xsep=12pt](r1) {};
        \draw[replace] (r1) to (f1);
      \end{scope}
    \end{scope}
    \begin{scope}[xshift=9cm,yshift=0.8cm]
      \node[ext] (n) at (0,0) { $\rv{N}_1$ };
      \node[var] (n1) at (2,0) { $\rv{W}_2$ };
      \node[fac] at (1,0) {} edge (n) edge (n1);
      \begin{scope}[on background layer]
        \node[fit=(n)(n1),fill=gray!20,inner xsep=12pt] (r2) {};
        \draw[replace] (r2) to (f4);
      \end{scope}
    \end{scope}
    \begin{scope}[xshift=9cm,yshift=-0.8cm]
      \node[ext] (n) at (0,0) { $\rv{N}_1$ };
      \node[var] (n1) at (2,0) { $\rv{W}_2$ };
      \node[fac] at (1,0) {} edge (n) edge (n1);
      \begin{scope}[on background layer]
        \node[fit=(n)(n1),fill=gray!20,inner xsep=12pt] (r3) {};
        \draw[replace] (r3) to (f5);
      \end{scope}
    \end{scope}
  \end{tikzpicture}
\end{center}}
\end{example}
We draw external nodes in black.
We draw an edge with nonterminal label $X$ as a square with $X$ inside.
Although the left-hand side is just a nonterminal symbol, we draw it like an edge, with replicas of the external nodes as attachment nodes.

The graphs generated by an FGG can be viewed, together with $D$ and $F$, as factor graphs, each of which defines a distribution over assignments.
We also want each nonterminal of the FGG to be thought of as a distribution over derivation trees and assignments; for present purposes, we only consider the marginal distribution over assignments to external nodes:
\begin{definition} \label{def:fggweight}
An \emph{assignment} of a nonterminal edge label~$X$ is a sequence $\asst \in D^{\arity(X)}$ of values.
The \emph{weight} of an assignment~$\asst$~is
\begin{equation}
    w_{X}(\asst) =
    \sum_{T \in \mathcal{D}(X)}
    w_{\ext_T}(\asst),
\end{equation}
where $\ext_T$ is the list of external nodes of the graph yielded by $T$.
\end{definition}

To query some variables, we can make them external in the generated graphs (because $S$ can have nonzero arity), and $w_S$ is their joint distribution. \Citet{chiang+riley:2020} discuss how to compute $w_S$; in particular, if all variables have finite domain, then $w_S$ can be computed in time linear in the size of the FGG and exponential in the maximum treewidth of its right-hand sides.

\section{A Simple Probabilistic Programming Language}

The source language of our translation has the following syntax.
Because the language is first-order, variable names~$x$ are distinct
from function names~$f$.
\begin{align*}
&\text{Programs} & q &\bnf e
    \bnfalt \textbf{fun}~f(x_1, \ldots, x_n) = e ; q \\
&\text{Expressions} & e &\bnf x
    \bnfalt \textbf{let}~x_1 = e_2~\textbf{in}~e_3
    \bnfalt f(e_1, \ldots, e_n)
    \bnfalt \textbf{sample}~e_1
    \bnfalt \textbf{observe}~e_1 \leftarrow e_2
    \\
&&& \bnfalt \textbf{if}~e_1~\textbf{then}~e_2~\textbf{else}~e_3
    \bnfalt \textbf{case}~e_1~\textbf{of}~\inl{x_2} \Rightarrow e_2 \alt \inr{x_3} \Rightarrow e_3
\end{align*}
We assume a number of built-in functions and constants: $\mathord=$, $\mathord{\neq}$, $\mathit{true}$, $\mathit{false}$, pairing, \textit{fst}, \textit{snd}, \textit{inl}, \textit{inr}, \textit{unit}. Boolean operations $\textbf{and}$ and $\textbf{or}$ can be treated as built-in functions, or if short-circuiting is desired, defined as syntactic sugar:
\begin{align*}
e_1~\textbf{and}~e_2 &\equiv \textbf{if}~e_1~\textbf{then}~e_2~\textbf{else}~\textit{false} &
e_1~\textbf{or}~e_2 &\equiv \textbf{if}~e_1~\textbf{then}~\textit{true}~\textbf{else}~e_2
\end{align*}

The only probabilistic constructs are \textbf{sample} and \textbf{observe}.
Generatively speaking,
$\textbf{sample}~e_1$ evaluates
$e_1$ to a distribution and then nondeterministically samples from the distribution,
whereas $\textbf{observe}~e_1 \leftarrow e_2$ evaluates
$e_2$ to a distribution and then
multiplies the weight of the current branch of computation
by the density of the distribution at the value of~$e_1$.
Other constructs can be defined as syntactic sugar in terms of $\textbf{observe}$,
by setting $e_2$ to a common distribution such as Bernoulli or exponential.
In particular, a~\textbf{fail} expression unconditionally terminates the current branch of computation, multiplying its weight by zero.

We assume that $D$ is countable, and define the denotations of the language in a relatively standard way.
Without recursion, we would directly define the denotation of each expression~$e$ to be an s-finite kernel from environments to values \citep{staton-commutative}, where an environment is a tuple that maps each free variable of~$e$ to its value.
With recursion, we have to first define this denotation relative to an interpretation of each function~$f$ as itself an s-finite kernel from environments to values, where an environment is a tuple that maps each argument of~$f$ to its value.
Thus, the denotation of a program's function definitions maps an interpretation of each function to an interpretation of each function.
Finally, we take the least fixed point of this monotonic map \citep{kozen-semantics}.

\section{Compiling Probabilistic Programs to FGGs}
\label{sec:ppl}

Every subexpression~$e$ or function~$f$ of a program can be translated into an HRG production (or two) whose left-hand-side nonterminal is $e$ or~$f$, respectively. 
If the subexpression~$e$ occurs in an environment with $k$ bound variables, or if the function~$f$ takes $k$ arguments, then we translate it to a nonterminal with arity $k+1$. The external nodes are $x_1, \ldots, x_k$ for the variables and $v$ for the result.
Such a nonterminal is pictured below.
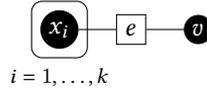
\begin{center}
\begin{tikzpicture}
\node[ext] (xi) {$x_i$};
\node[plate,fit=(xi),label=below:{$i=1, \ldots, k$}] {};
\node[fac,right=of xi](f) {$e$} edge (xi);
\node[ext,right=of f](v) {$v$} edge (f);
\end{tikzpicture}
\end{center}
We use plate notation to depict multiple nodes. Unless otherwise indicated, each plate has an instance for each $i=1, \ldots, k$. We stress, however, that plates are only meta-notation; as will be clear in the example below, actual FGG rules do \emph{not} use plates.

Our translation preserves semantics in the sense that the denotation of
a subexpression~$e$ or function~$f$, regarded as a weight function on
$(k+1)$-tuples, is equal to the weight function $w_e$ or~$w_f$ of its
translation (Definition~\ref{def:fggweight}).

\subsection{Random variables}

The \textbf{sample} and \textbf{observe} expressions are translated
by turning the density of the distribution into a factor:
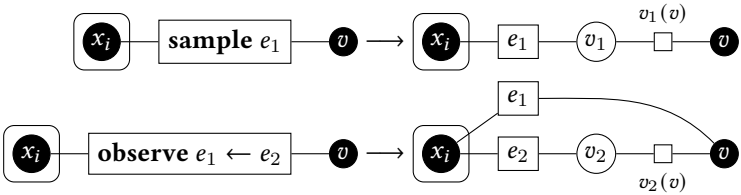
\begin{align*}
\begin{tikzpicture}
\node[ext](xi) {$x_i$};
\node[plate,fit=(xi)] {};
\node[fac,right=of xi](f) {$\textbf{sample}~e_1$} edge (xi);
\node[ext,right=of f](v) {$v$} edge (f);
\end{tikzpicture}
&\longrightarrow
\begin{tikzpicture}
\node[ext](xi) {$x_i$};
\node[plate,fit=(xi)] {};
\node[fac,right=of xi](f1) {$e_1$} edge (xi);
\node[var,right=of f1](v1) {$v_1$} edge (f1);
\node[fac,right=of v1,label={[overlay]above:$v_1(v)$}](f2) {} edge (v1);
\node[ext,right=of f2](v) {$v$} edge (f2);
\end{tikzpicture} \\
\begin{tikzpicture}
\node[ext](xi) {$x_i$};
\node[plate,fit=(xi)] {};
\node[fac,right=of xi](f) {$\textbf{observe}~e_1 \leftarrow e_2$} edge (xi);
\node[ext,right=of f](v) {$v$} edge (f);
\end{tikzpicture}
&\longrightarrow
\begin{tikzpicture}
\node[ext](xi) {$x_i$};
\node[plate,fit=(xi)] {};
\node[fac,right=of xi](f1) {$e_2$} edge (xi);
\node[var,right=of f1](v1) {$v_2$} edge (f1);
\node[fac,right=of v1,label=below:{$v_2(v)$}](f3) {} edge (v1);
\node[ext,right=of f3](v) {$v$} edge (f3);
\node[fac,above right=of xi.east](f2) {$e_1$} edge (xi) edge[out=0,in=135] (v);
\end{tikzpicture}
\end{align*}

\subsection{Conditionals and disjoint unions}

A conditional expression translates to \emph{two} productions, one for the case where the condition is true and one for the case where the condition is false:
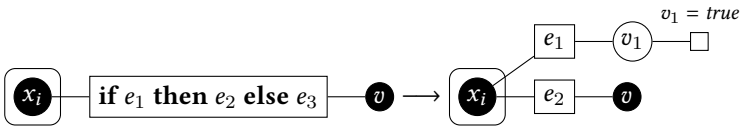
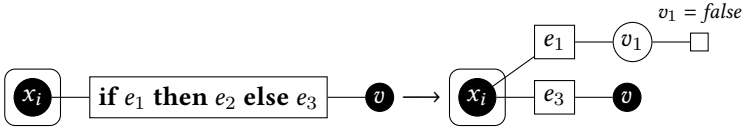
\begin{align*}
\begin{tikzpicture}
\node[ext](xi) {$x_i$};
\node[plate,fit=(xi)] {};
\node[fac,right=of xi](f) {$\textbf{if}~e_1~\textbf{then}~e_2~\textbf{else}~e_3$} edge (xi);
\node[ext,right=of f](v) {$v$} edge (f);
\end{tikzpicture}
&\longrightarrow
\begin{tikzpicture}
\node[ext](xi) {$x_i$};
\node[plate,fit=(xi)] {};
\node[fac,above right=of xi.east](f1) {$e_1$} edge (xi);
\node[var,right=of f1](v1) {$v_1$} edge (f1);
\node[fac,right=of v1,label=above:{$v_1=\textit{true}$}] {} edge (v1);
\node[fac,right=of xi](f2) {$e_2$} edge (xi);
\node[ext,right=of f2](v2) {$v$} edge (f2);
\end{tikzpicture} \\
\begin{tikzpicture}
\node[ext](xi) {$x_i$};
\node[plate,fit=(xi)] {};
\node[fac,right=of xi](f) {$\textbf{if}~e_1~\textbf{then}~e_2~\textbf{else}~e_3$} edge (xi);
\node[ext,right=of f](v) {$v$} edge (f);
\end{tikzpicture}
&\longrightarrow
\begin{tikzpicture}
\node[ext](xi) {$x_i$};
\node[plate,fit=(xi)] {};
\node[fac,above right=of xi.east](f1) {$e_1$} edge (xi);
\node[var,right=of f1](v1) {$v_1$} edge (f1);
\node[fac,right=of v1,label=above:{$v_1=\textit{false}$}] {} edge (v1);
\node[fac,right=of xi](f2) {$e_3$} edge (xi);
\node[ext,right=of f2](v2) {$v$} edge (f2);
\end{tikzpicture}
\end{align*}
This is the only place a left-hand side has more than one right-hand side (along with the related translation of $\textbf{case}$). So the FGG has one derivation tree for each possible code path.

This translation avoids a problem that \citet[Section 3.1]{vandemeent+:2018} encounter when translating conditional expressions to factor graphs. In a factor graph, both arms of the conditional must be translated, and every assignment to the factor graph must assign values to variables in both arms, even though only one can be active at a time. Their translation requires some complicated machinery to work around this problem. But our translation to an FGG does not have this problem, because it generates a different graph for each arm.

The translation of $\textbf{case}$ is similar to $\textbf{if}$:
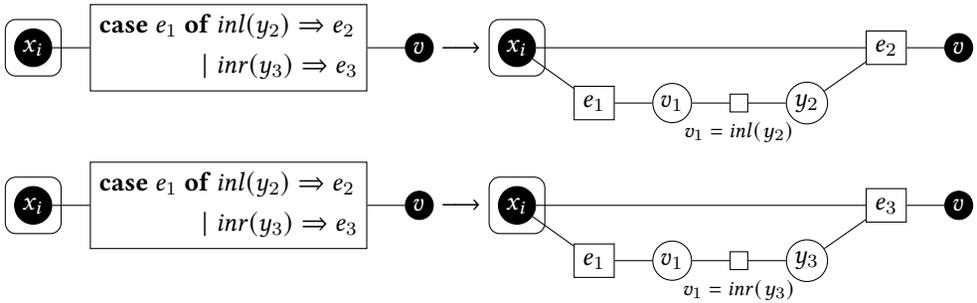
\begin{align*}
\begin{tikzpicture}
\node[ext](xi) {$x_i$};
\node[plate,fit=(xi)] {};
\node[fac,right=of xi](f) {$\begin{aligned}\textbf{case}~e_1~\textbf{of}~&\inl{y_2} \Rightarrow e_2 \\ \alt {} &\inr{y_3} \Rightarrow e_3\end{aligned}$} edge (xi);
\node[ext,right=of f](v) {$v$} edge (f);
\end{tikzpicture}
&\longrightarrow
\begin{tikzpicture}
\node[ext](xi) {$x_i$};
\node[plate,fit=(xi)] {};
\node[fac,below right=of xi.east](f1) {$e_1$} edge (xi);
\node[var,right=of f1](v1) {$v_1$} edge (f1);
\node[fac,right=of v1,label=below:{$v_1=\inl{y_2}$}](f1a) {} edge (v1);
\node[var,right=of f1a](y2) {$y_2$} edge (f1a);
\node[fac,above right=of y2.east](f2) {$e_2$} edge (xi) edge (y2);
\node[ext,right=of f2](v2) {$v$} edge (f2);
\end{tikzpicture} \\
\begin{tikzpicture}
\node[ext](xi) {$x_i$};
\node[plate,fit=(xi)] {};
\node[fac,right=of xi](f) {$\begin{aligned}\textbf{case}~e_1~\textbf{of}~&\inl{y_2} \Rightarrow e_2 \\ \alt {} &\inr{y_3} \Rightarrow e_3\end{aligned}$} edge (xi);
\node[ext,right=of f](v) {$v$} edge (f);
\end{tikzpicture}
&\longrightarrow
\begin{tikzpicture}
\node[ext](xi) {$x_i$};
\node[plate,fit=(xi)] {};
\node[fac,below right=of xi.east](f1) {$e_1$} edge (xi);
\node[var,right=of f1](v1) {$v_1$} edge (f1);
\node[fac,right=of v1,label=below:{$v_1=\inr{y_3}$}](f1a) {} edge (v1);
\node[var,right=of f1a](y2) {$y_3$} edge (f1a);
\node[fac,above right=of y2.east](f2) {$e_3$} edge (xi) edge (y2);
\node[ext,right=of f2](v2) {$v$} edge (f2);
\end{tikzpicture}
\end{align*}

\subsection{Functions and local variables}

A function definition $\textbf{fun}~f(x_1, \ldots, x_n) = e$ becomes the production:
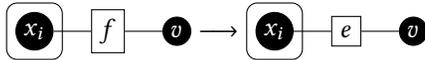
\begin{align*}
\begin{tikzpicture}
\node[ext](xi) {$x_i$};
\node[plate,fit=(xi)] {};
\node[fac,right=of xi](f) {$f$} edge (xi);
\node[ext,right=of f](v) {$v$} edge (f);
\end{tikzpicture}
&\longrightarrow
\begin{tikzpicture}
\node[ext](xi) {$x_i$};
\node[plate,fit=(xi)] {};
\node[fac,right=of xi](f) {$e$} edge (xi);
\node[ext,right=of f](v) {$v$} edge (f);
\end{tikzpicture}
\\
\intertext{Inside the function body, variables evaluate to their value in the environment:}
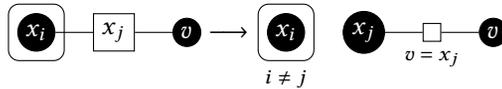
\begin{tikzpicture}
\node[ext](xi) {$x_i$};
\node[plate,fit=(xi)] {};
\node[fac,right=of xi](f) {$x_j$} edge (xi);
\node[ext,right=of f](v) {$v$} edge (f);
\end{tikzpicture}
&\longrightarrow
\begin{tikzpicture}
\node[ext](xi) {$x_i$};
\node[ext,right=of xi](xj) {$x_j$};
\node[plate,fit=(xi),label=below:{$i\neq j$}] {};
\node[fac,right=of xj,label=below:{$v=x_j$}](f) {} edge (xj);
\node[ext,right=of f](v) {$v$} edge (f);
\end{tikzpicture} \\
\intertext{And function calls become:}
\begin{tikzpicture}
\node[ext](xi) {$x_i$};
\node[plate,fit=(xi)] {};
\node[fac,right=of xi](f) {$f(e_1, \ldots, e_n)$} edge (xi);
\node[ext,right=of f](v) {$v$} edge (f);
\end{tikzpicture}
&\longrightarrow
\begin{tikzpicture}
\node[ext](xi) {$x_i$};
\node[plate,fit=(xi)] {};
\node[fac,right=of xi](f1) {$e_j$} edge (xi);
\node[var,right=of f1](v1) {$v_j$} edge (f1);
\node[plate,fit=(f1)(v1),label=below:{$j=1, \ldots, n$}] {};
\node[fac,right=of v1](f) {$f$} edge (v1);
\node[ext,right=of f](v) {$v$} edge (f);
\end{tikzpicture}
\end{align*}
Unlike in the translation of \citet{vandemeent+:2018}, the function can be recursive: the definition of $f$ can include calls to $f$. Infinite recursion is disallowed because the definition of FGG doesn't allow infinite-sized derivation trees. But a program can have an infinite number of branches of computation, in which the depth of recursion is unbounded. The resulting FGG has an infinite number of derivation trees, and the inference algorithm does sum over, or at least approximate the sum over, all of them.

The translation of $\textbf{let}$ is:
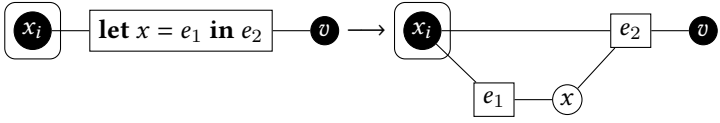
\begin{align*}
\begin{tikzpicture}
\node[ext](xi) {$x_i$};
\node[plate,fit=(xi)] {};
\node[fac,right=of xi](f) {$\textbf{let}~x=e_1~\textbf{in}~e_2$} edge (xi);
\node[ext,right=of f](v) {$v$} edge (f);
\end{tikzpicture}
&\longrightarrow
\begin{tikzpicture}
\node[ext](xi) {$x_i$};
\node[plate,fit=(xi)] {};
\node[fac,below right=of xi](f1) {$e_1$} edge (xi);
\node[var,right=of f1](v1) {$x$} edge (f1);
\node[right=of f1.east](c) {$\phantom{e_2}$};
\node[fac,right=of v1](f2) at (xi-|c) {$e_2$} edge (v1) edge (xi);
\node[ext,right=of f2](v) {$v$} edge (f2);
\end{tikzpicture}
\end{align*}

\subsection{Built-in functions and constants}

Built-in functions and constants
\begin{align*}
  c &\bnf \mathord{=} \bnfalt \mathord{\neq}
    \bnfalt \textit{true} \bnfalt \textit{false}
    \bnfalt (,) \bnfalt \textit{fst} \bnfalt \textit{snd}
    \bnfalt \textit{inl} \bnfalt \textit{inr}
    \bnfalt \textit{unit}
\end{align*}
can be translated into FGG rules that give rise to terminal edges:
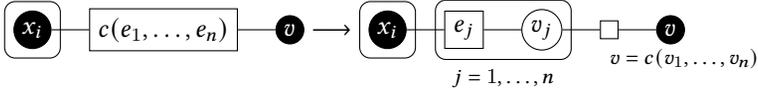
\begin{align*}
\begin{tikzpicture}
\node[ext](xi) {$x_i$};
\node[plate,fit=(xi)] {};
\node[fac,right=of xi](f) {$c(e_1, \ldots, e_n)$} edge (xi);
\node[ext,right=of f](v) {$v$} edge (f);
\end{tikzpicture}
&\longrightarrow
\begin{tikzpicture}
\node[ext](xi) {$x_i$};
\node[plate,fit=(xi)] {};
\node[fac,right=of xi](f1) {$e_j$} edge (xi);
\node[var,right=of f1](v1) {$v_j$} edge (f1);
\node[plate,fit=(f1)(v1),label=below:{$j=1, \ldots, n$}] {};
\node[fac,right=of v1,label={[xshift=1cm]below:$v=c(v_1, \ldots, v_n)$}](f) {} edge (v1);
\node[ext,right=of f](v) {$v$} edge (f);
\end{tikzpicture}
\end{align*}

\subsection{Programs}

Finally, if $e$ is the top-level expression at the end of a program ($q$), create the rule
\begin{align*}
\begin{tikzpicture}
\node[fac](s) {$S$};
\node[ext,right=of s] {$v$} edge (s);
\end{tikzpicture}
&\longrightarrow
\begin{tikzpicture}
\node[fac](e) {$e$};
\node[ext,right=of e] {$v$} edge (e);
\end{tikzpicture}
\end{align*}
where $S$ is the start symbol of the FGG\@.

\section{Example: PCFG}

The code in Figure~\ref{fig:code}a nondeterministically samples a string from a PCFG in Chomsky normal form. If $X$ is a left-hand side, let $p[X]$ be a distribution over right-hand sides, which can be $\inl{a}$ for a single terminal symbol $a$ or $\inr{Y, Z}$ for two nonterminals $Y$ and $Z$.

This translates to the FGG in Figure~\ref{fig:fgg_pcfg}; note the similarity to Example~\ref{eg:fgg_pcfg}. We have made a few optimizations for greater readability:
\begin{itemize}
\item We ``inline'' all rules except those for \textbf{if}, \textbf{case}, and functions.
\item Consecutive terminal edges arising from built-in functions and constants are composed into a single terminal edge.
\item In the translation of variables, the factor $v = x_j$ can usually be contracted, merging the nodes for $x_j$ and $v$.
\item Rules that must have zero weight are omitted.
\end{itemize}

\begin{figure}
\begin{tabular}{cc}
\begin{minipage}{2in}
\begin{verbatim}
fun d(x) =
  case sample p[x] of
    inl a =>
      unit
  | inr yz =>
      let u = d(fst(yz)) in
      d(snd(yz));
d(S)
\end{verbatim}
\end{minipage}
    &
\begin{minipage}{3in}
\begin{verbatim}
fun d(w, x) =
  case sample p[x] of
    inl a =>
      if w != nil and car(w) = a then
        cdr(w)
      else
        fail
  | inr yz =>
      let w' = d(w, fst(yz)) in
      d(w', snd(yz));
if d(w, S) = nil then unit else fail
\end{verbatim}
\end{minipage}
\\
\\
(a) & (b)
\end{tabular}
\caption{Code for (a) all derivations of a PCFG in Chomsky normal form; (b) all derivations yielding a given string $w$.}
\label{fig:code}

\begin{center}
  \shrink{\begin{align*}
\begin{tikzpicture}[node distance=1cm,x=1cm]
\node[fac](e) at (0,-0.0) {$S$};
\node[ext,right of=e](v1) {$v_{1}$};
\draw (e) edge[out=0,in=180] (v1);
\end{tikzpicture}
&\longrightarrow
\begin{tikzpicture}[node distance=1cm,x=1cm]
\node[fac,label={[text width=1.9cm,align=center]below:$v_{2}=S$}](e2) at (0,-0.0) {};
\node[var,right of=e2](v2) {$v_{2}$};
\node[fac,right of=v2](e1) {$d$};
\node[ext,right of=e1](v1) {$v_{1}$};
\draw (e2) edge[out=0,in=180] (v2);
\draw (e1) edge[out=0,in=180] (v1);
\draw (v2) edge[out=0,in=180] (e1);
\end{tikzpicture}
\\
\begin{tikzpicture}[node distance=1cm,x=1cm]
\node[ext](x) at (0,-0.0) {$x$};
\node[fac,right of=x](e3) {$d$};
\node[ext,right of=e3](v3) {$v_{7}$};
\draw (x) edge[out=0,in=180] (e3);
\draw (e3) edge[out=0,in=180] (v3);
\end{tikzpicture}
&\longrightarrow
\begin{tikzpicture}[node distance=1cm,x=1cm]
\node[ext](x) at (0,-0.0) {$x$};
\node[fac,label={[text width=1.9cm,align=center]below:$v_{5}=p[x]$},right of=x](e5) {};
\node[var,right of=e5](v5) {$v_{5}$};
\node[fac,label={[text width=1.9cm,align=center]below:$v_{5}(v_{4})$},right of=v5](e4) {};
\node[var,right of=e4](v4) {$v_{4}$};
\node[fac,label={[text width=1.9cm,align=center]below:$v_{4}=\inl{a}$},right of=v4](v3_l) {};
\node[var,right of=v3_l](a) {$a$};
\node[fac,label={[text width=1.9cm,align=center]below:$v_{7}=\mathit{unit}$}](e7) at (7,-0.0) {};
\node[ext,right of=e7](v7) {$v_{7}$};
\draw (v5) edge[out=0,in=180] (e4);
\draw (v4) edge[out=0,in=180] (v3_l);
\draw (e7) edge[out=0,in=180] (v7);
\draw (v3_l) edge[out=0,in=180] (a);
\draw (e5) edge[out=0,in=180] (v5);
\draw (x) edge[out=0,in=180] (e5);
\draw (e4) edge[out=0,in=180] (v4);
\end{tikzpicture}
\\
\begin{tikzpicture}[node distance=1cm,x=1cm]
\node[ext](x) at (0,-0.0) {$x$};
\node[fac,right of=x](e3) {$d$};
\node[ext,right of=e3](v3) {$v_{12}$};
\draw (x) edge[out=0,in=180] (e3);
\draw (e3) edge[out=0,in=180] (v3);
\end{tikzpicture}
&\longrightarrow
\begin{tikzpicture}[node distance=1cm,x=1cm]
\node[ext](x) at (0,-0.0) {$x$};
\node[fac,label={[text width=1.9cm,align=center]below:$v_{5}=p[x]$},right of=x](e5) {};
\node[var,right of=e5](v5) {$v_{5}$};
\node[fac,label={[text width=1.9cm,align=center]below:$v_{5}(v_{4})$},right of=v5](e4) {};
\node[var,right of=e4](v4) {$v_{4}$};
\node[fac,label={[text width=1.9cm,align=center]below:$v_{4}=\inr{yz}$},right of=v4](v3_r) {};
\node[var,right of=v3_r](yz) {$yz$};
\node[fac,label={[text width=1.9cm,align=center]above:$v_{10}=\mathit{fst\/}(yz)$}](e10) at (7,0.5) {};
\node[fac,label={[text width=1.9cm,align=center]below:$v_{13}=\mathit{snd\/}(yz)$}](e13) at (7,-0.5) {};
\node[var,right of=e10](v10) {$v_{10}$};
\node[var,right of=e13](v13) {$v_{13}$};
\node[fac,right of=v10](e9) {$d$};
\node[fac,right of=v13](e12) {$d$};
\node[var,right of=e9](v9) {$u$};
\node[ext,right of=e12](v12) {$v_{12}$};
\draw (v5) edge[out=0,in=180] (e4);
\draw (yz) edge[out=0,in=180] (e13);
\draw (v3_r) edge[out=0,in=180] (yz);
\draw (e13) edge[out=0,in=180] (v13);
\draw (v4) edge[out=0,in=180] (v3_r);
\draw (e9) edge[out=0,in=180] (v9);
\draw (e10) edge[out=0,in=180] (v10);
\draw (v10) edge[out=0,in=180] (e9);
\draw (e5) edge[out=0,in=180] (v5);
\draw (x) edge[out=0,in=180] (e5);
\draw (e12) edge[out=0,in=180] (v12);
\draw (e4) edge[out=0,in=180] (v4);
\draw (v13) edge[out=0,in=180] (e12);
\draw (yz) edge[out=0,in=180] (e10);
\end{tikzpicture}
\end{align*}}
\end{center}
\caption{Translation of the program in Figure~\ref{fig:code}a.}
\label{fig:fgg_pcfg}

\begin{center}
  \shrink{\begin{align*}
\begin{tikzpicture}[node distance=1cm,x=1cm]
\node[fac](e1) at (0,-0.0) {$S$};
\node[ext,right of=e1](v1) {$v_{6}$};
\draw (e1) edge[out=0,in=180] (v1);
\end{tikzpicture}
&\longrightarrow
\begin{tikzpicture}[node distance=1cm,x=1cm]
\node[fac,label={[text width=1.9cm,align=center]below:$v_{4}=w$}](e4) at (0,1.0) {};
\node[fac,label={[text width=1.9cm,align=center]below:$v_{5}=S$}](e5) at (0,0.0) {};
\node[fac,label={[text width=1.9cm,align=center]below:$v_{6}=\mathit{unit}$}](e6) at (0,-1.0) {};
\node[var,right of=e4](v4) {$v_{4}$};
\node[var,right of=e5](v5) {$v_{5}$};
\node[ext,right of=e6](v6) {$v_{6}$};
\node[fac](e3) at (2,0.0) {$d$};
\node[var,right of=e3](v3) {$v_{3}$};
\node[fac,label={[text width=1.9cm,align=center]below:$v_{2}=v_{3}=\mathit{nil}$},right of=v3](e2) {};
\node[var,right of=e2](v2) {$v_{2}$};
\node[fac,label={[text width=1.9cm,align=center]below:$v_{2}=\textit{true}$},right of=v2](v1_t) {};
\draw (v4) edge[out=0,in=180] (e3);
\draw (v2) edge[out=0,in=180] (v1_t);
\draw (v3) edge[out=0,in=180] (e2);
\draw (e5) edge[out=0,in=180] (v5);
\draw (e4) edge[out=0,in=180] (v4);
\draw (e6) edge[out=0,in=180] (v6);
\draw (e2) edge[out=0,in=180] (v2);
\draw (v5) edge[out=0,in=180] (e3);
\draw (e3) edge[out=0,in=180] (v3);
\end{tikzpicture}
\\
\begin{tikzpicture}[node distance=1cm,x=1cm]
\node[ext](w) at (0,0.5) {$w$};
\node[ext](x) at (0,-0.5) {$x$};
\node[fac](e8) at (1,-0.0) {$d$};
\node[ext,right of=e8](v8) {$v_{17}$};
\draw (e8) edge[out=0,in=180] (v8);
\draw (x) edge[out=0,in=180] (e8);
\draw (w) edge[out=0,in=180] (e8);
\end{tikzpicture}
&\longrightarrow
\begin{tikzpicture}[node distance=1cm,x=1cm]
\node[ext](w) at (0,1.0) {$w$};
\node[ext](x) at (0,-0.0) {$x$};
\node[coordinate,right of=w](virtual_1) {$$};
\node[fac,label={[text width=1.9cm,align=center]below:$v_{10}=p[x]$},right of=x](e10) {};
\node[fac,label={[text width=1.9cm,align=center]below:$v_{17}=\mathit{cdr\/}(w)$}](e17) at (1,-1.0) {};
\node[coordinate,right of=virtual_1](virtual_2) {$$};
\node[var,right of=e10](v10) {$v_{10}$};
\node[ext,right of=e17](v17) {$v_{17}$};
\node[coordinate,right of=virtual_2](virtual_3) {$$};
\node[fac,label={[text width=1.9cm,align=center]below:$v_{10}(v_{9})$},right of=v10](e9) {};
\node[coordinate,right of=virtual_3](virtual_4) {$$};
\node[var,right of=e9](v9) {$v_{9}$};
\node[coordinate,right of=virtual_4](virtual_5) {$$};
\node[fac,label={[text width=1.9cm,align=center]below:$v_{9}=\inl{a}$},right of=v9](v8_l) {};
\node[coordinate,right of=virtual_5](virtual_6) {$$};
\node[var,right of=v8_l](a) {$a$};
\node[fac,label={[text width=1.9cm,align=center]below:$v_{13}=w\neq \mathit{nil}\land \mathit{car\/}(w)=a$}](e13) at (7,0.0) {};
\node[var,right of=e13](v13) {$v_{13}$};
\node[fac,label={[text width=1.9cm,align=center]below:$v_{13}=\textit{true}$},right of=v13](v12_t) {};
\draw (virtual_3) edge[out=0,in=180] (virtual_4);
\draw (virtual_2) edge[out=0,in=180] (virtual_3);
\draw (v8_l) edge[out=0,in=180] (a);
\draw (w) edge[out=0,in=180] (virtual_1);
\draw (e13) edge[out=0,in=180] (v13);
\draw (virtual_5) edge[out=0,in=180] (virtual_6);
\draw (e17) edge[out=0,in=180] (v17);
\draw (v13) edge[out=0,in=180] (v12_t);
\draw (e10) edge[out=0,in=180] (v10);
\draw (e9) edge[out=0,in=180] (v9);
\draw (v10) edge[out=0,in=180] (e9);
\draw (x) edge[out=0,in=180] (e10);
\draw (a) edge[out=0,in=180] (e13);
\draw (v9) edge[out=0,in=180] (v8_l);
\draw (virtual_1) edge[out=0,in=180] (virtual_2);
\draw (virtual_6) edge[out=0,in=180] (e13);
\draw (virtual_4) edge[out=0,in=180] (virtual_5);
\draw (w) edge[out=0,in=180] (e17);
\end{tikzpicture}
\\
\begin{tikzpicture}[node distance=1cm,x=1cm]
\node[ext](w) at (0,0.5) {$w$};
\node[ext](x) at (0,-0.5) {$x$};
\node[fac](e8) at (1,-0.0) {$d$};
\node[ext,right of=e8](v8) {$v_{25}$};
\draw (e8) edge[out=0,in=180] (v8);
\draw (x) edge[out=0,in=180] (e8);
\draw (w) edge[out=0,in=180] (e8);
\end{tikzpicture}
&\longrightarrow
\begin{tikzpicture}[node distance=1cm,x=1cm]
\node[ext](w) at (0,1.0) {$w$};
\node[ext](x) at (0,-0.5) {$x$};
\node[coordinate,right of=w](virtual_1) {$$};
\node[fac,label={[text width=1.9cm,align=center]below:$v_{10}=p[x]$},right of=x](e10) {};
\node[coordinate,right of=virtual_1](virtual_2) {$$};
\node[var,right of=e10](v10) {$v_{10}$};
\node[coordinate,right of=virtual_2](virtual_3) {$$};
\node[fac,label={[text width=1.9cm,align=center]below:$v_{10}(v_{9})$},right of=v10](e9) {};
\node[coordinate,right of=virtual_3](virtual_4) {$$};
\node[var,right of=e9](v9) {$v_{9}$};
\node[coordinate,right of=virtual_4](virtual_5) {$$};
\node[fac,label={[text width=1.9cm,align=center]below:$v_{9}=\inr{yz}$},right of=v9](v8_r) {};
\node[coordinate,right of=virtual_5](virtual_6) {$$};
\node[var,right of=v8_r](yz) {$yz$};
\node[coordinate,right of=virtual_6](virtual_7) {$$};
\node[fac,label={[text width=1.9cm,align=center]above:$v_{23}=\mathit{fst\/}(yz)$}](e23) at (7,-0.0) {};
\node[fac,label={[text width=1.9cm,align=center]below:$v_{27}=\mathit{snd\/}(yz)$}](e27) at (7,-1.0) {};
\node[coordinate,right of=virtual_7](virtual_8) {$$};
\node[var,right of=e23](v23) {$v_{23}$};
\node[var,right of=e27](v27) {$v_{27}$};
\node[fac](e21) at (9,-0.0) {$d$};
\node[coordinate,right of=v27](virtual_22) {$$};
\node[var,right of=e21](v21) {$w'$};
\node[coordinate,right of=virtual_22](virtual_23) {$$};
\node[fac](e25) at (11,-0.5) {$d$};
\node[ext,right of=e25](v25) {$v_{25}$};
\draw (virtual_22) edge[out=0,in=180] (virtual_23);
\draw (yz) edge[out=0,in=180] (e27);
\draw (virtual_5) edge[out=0,in=180] (virtual_6);
\draw (virtual_7) edge[out=0,in=180] (virtual_8);
\draw (e9) edge[out=0,in=180] (v9);
\draw (x) edge[out=0,in=180] (e10);
\draw (virtual_4) edge[out=0,in=180] (virtual_5);
\draw (e21) edge[out=0,in=180] (v21);
\draw (virtual_6) edge[out=0,in=180] (virtual_7);
\draw (v27) edge[out=0,in=180] (virtual_22);
\draw (v10) edge[out=0,in=180] (e9);
\draw (v23) edge[out=0,in=180] (e21);
\draw (e23) edge[out=0,in=180] (v23);
\draw (yz) edge[out=0,in=180] (e23);
\draw (virtual_1) edge[out=0,in=180] (virtual_2);
\draw (virtual_8) edge[out=0,in=180] (e21);
\draw (e27) edge[out=0,in=180] (v27);
\draw (v9) edge[out=0,in=180] (v8_r);
\draw (virtual_2) edge[out=0,in=180] (virtual_3);
\draw (e25) edge[out=0,in=180] (v25);
\draw (virtual_23) edge[out=0,in=180] (e25);
\draw (virtual_3) edge[out=0,in=180] (virtual_4);
\draw (w) edge[out=0,in=180] (virtual_1);
\draw (v8_r) edge[out=0,in=180] (yz);
\draw (e10) edge[out=0,in=180] (v10);
\draw (v21) edge[out=0,in=180] (e25);
\end{tikzpicture}
\end{align*}}
\end{center}
\caption{Translation of the program in Figure~\ref{fig:code}b.}
\label{fig:fgg_pcfgw}
\end{figure}

Next, we modify this program to take a string $w$ as input and constrain the derivations to those that yield $w$, shown in Figure~\ref{fig:code}b. This translates to the FGG in Figure~\ref{fig:fgg_pcfgw}.
Computing $w_S$ on this grammar \citep{chiang+riley:2020} is equivalent to CKY\@. In the last rule, the nodes $w$, $w'$, and $v_{25}$ range over suffixes of the input string, so summing over assignments to the right-hand side takes $O(n^3)$ time, just as in CKY\@.
Unlike a typical implementation of CKY\@, however, the programs in Figure~\ref{fig:code} can be easily modified to efficiently handle, say, the intersection of a PCFG with a regular language \citep{lang-1988-parsing,hale-2004-information}, or a PCFG whose nonterminals are structured as tuples \citep{collins-three,klein-accurate}.

\section{Conclusion}

The above translation has been implemented, but only outputs \LaTeX{} code to produce the diagrams shown in this paper. We plan to properly implement the translation to output an FGG, but first we need to implement FGGs and inference algorithms for them.

The translation only produces FGGs in which each rule has exactly one output external node. As far as FGGs are concerned, a rule can have more than one output external node; the nonterminal edge that ``calls'' it could even have an input node that depends on one of its output nodes. What kinds of functions would translate into such rules?

\begin{acks}
This material is based upon work supported by the \grantsponsor{GS100000001}{National Science Foundation}{http://dx.doi.org/10.13039/100000001} under Grant Nos.~\grantnum{GS100000001}{2019291} and~\grantnum{GS100000001}{2019266}.  Any opinions, findings, and conclusions or recommendations expressed in this material are those of the authors and do not necessarily reflect the views of the National Science Foundation.
\end{acks}

\bibliography{references}

\end{document}